\documentclass[twocolumn,showpacs,preprintnumbers,amsmath,amssymb,10pt,a4paper]{revtex4}

\usepackage{geometry}
\usepackage{graphicx}% Include figure files
\usepackage{dcolumn}% Align table columns on decimal point
\usepackage{bm}% bold math
\newcommand{\be}{\begin{equation}}
\newcommand{\ee}{\end{equation}}
\newcommand{\bd}{\begin{displaymath}}
\newcommand{\ed}{\end{displaymath}}
\newcommand{\BE}{\begin{eqnarray}}
\newcommand{\EE}{\end{eqnarray}}

\newcommand{\bk}{\ensuremath{\mathbf{k}}}

\newcommand{\bx}{\ensuremath{\mathbf{x}}}

\newcommand{\bn}{\ensuremath{\mathbf{n}}}

\newcommand{\boldpsi}{{\mbox{\boldmath $\psi$}}}

\newcommand{\avg}[1]{\left\langle{#1}\right\rangle}

\begin{document}

\preprint{}
\title{Birth-death processes with quenched uncertainty and intrinsic noise}
% Force line breaks with \\

\author{Tobias Galla}
\email{tobias.galla@manchester.ac.uk}

\affiliation{Theoretical Physics, School of Physics and Astronomy, The University of Manchester, Manchester M13 9PL, United Kingdom}

\date{\today}% It is always \today, today,
             %  but any date may be explicitly specified

\begin{abstract}
The dynamics of populations is frequently subject to intrinsic noise. At the same time unknown interaction networks or rate constants can present quenched uncertainty.  Existing approaches often involve repeated sampling of the quenched disorder and then running the stochastic birth-death dynamics on these samples.  In this paper we take a different view, and formulate an effective jump process, representative of the ensemble of quenched interactions as a whole. Using evolutionary games with random payoff matrices as an example, we develop an algorithm to simulate this process, and we discuss diffusion approximations in the limit of weak intrinsic noise. 
\end{abstract}

\pacs{87.10.Mn, 02.50.Ey, 05.10.Gg, 87.23.Kg}% PACS, the Physics and Astronomy
                             % Classification Scheme.
%\keywords{Suggested keywords}%Use showkeys class option if keyword
                              %display desired
\maketitle
%%%%%%%%%%%%%%%%%%%%%%%%%%%%%%%%%%%%%%%%%%%%%%%%%%%%%%%%%%%%%%%%%%%%%%%%%%%%%%%%
%%%%%%%%%%%%%%%%%%%%%%%%%%%%%%%%%%%%%%%%%%%%%%%%%%%%%%%%%%%%%%%%%%%%%%%%%%%%%%%%
{\em Introduction.} The dynamics of populations frequently involve randomness. One type of noise is known as demographic or intrinsic stochasticity \cite{lande}. It results from the assumption that the births or deaths of individuals are random events \cite{melbourne}. This approach is commonly taken in models of population dynamics. Two identical members of the population with the same reproduction and death rates may generate different numbers of offspring, or die at different points in time. This variation is due to effects not described in more detail by the model; the finer details have been `integrated out' and only remain in the form of intrinsic noise. In addition to this, the parameters setting the birth and death rates, or the topology of the interaction network within the population may be unknown. This is a separate source of disorder, and leads to an elusive interplay of intrinsic dynamic noise and quenched extrinsic uncertainty. Understanding how different types of noise act together, and how uncertainty propagates is instrumental for many applications involving multi-scale models. It constitutes one of the main questions in the area of uncertainty quantification \cite{uq}. Numerical simulations frequently proceed by first drawing the reaction network or the rate constants from a distribution. This distribution is chosen to capture the believed uncertainty about the network's true structure or the actual values of the rate constants. In a second step one then runs the stochastic population dynamics on this quenched realisation of the interaction parameters. Statistics are collected by repeating the process for different random samples of the network or of the rate constants. 

Here, we take a different route. We ask whether it is possible to simulate one {\em effective} population dynamics, retaining intrinsic noise, but representative of the ensemble of the quenched interactions as a whole. To construct an answer for this problem we use a relatively simple birth-death process as an illustration. Our example is relevant for a variety problems, including populations interacting in games with random payoff matrices \cite{weini, multiverse, haigh1, haigh2, berg, broom, eriksson, cannings, han}, the evolution in random fitness landscapes \cite{krug1,krug2}, meta-populations and dynamics on random networks \cite{newman, rozhnova}, and  condensation phenomena in quantum systems which can also be described as birth-death processes \cite{knebel}. With appropriate extension, we believe that our answer to the opening question can be applied more broadly to discrete populations involving both intrinsic noise and quenched uncertainty. Our approach is based on a combination of tools from statistical physics. These include a path-integral approach to deal with the quenched disorder \cite{coolen, sollich, msr}, and a technique traditionally used to simulate the dynamical mean field theory of spin glasses and neural networks \cite{eissopp1, eissopp2}
\\
{\em Model.} We develop the method for populations of discrete individuals, who can each be of one of $S$ different species, labelled $i=1,\dots,S$. The population size is $N=S\times\Omega$. The parameter $\Omega$ is the scale of  the initial number of individuals per species. Our mathematical analysis applies in the simultaneous limits $S\to\infty$ and $N\to\infty$, but keeping the ratio $\Omega=N/S$ finite. The quantity $\Omega^{-1/2}$ sets the strength of the demographic noise. We write $n_i$ for the number of individuals of type $i$, and $\bn=(n_1,\dots,n_S)$. The continuous-time Markov process occurs through discrete birth-death events; in each of these events an individual of one type $i$ is removed from the population, and it is replaced by an individual of type $j$. The notation $T_{i\to j}(\bn)$ indicates the reaction rate for such an event. In our model they are of the form
\be\label{eq:rates}
T_{i\to j}=\frac{n_i n_j}{N}g(f_j, f_i),
\ee
where $f_i$ and $f_j$ characterise the reproductive fitnesses of species $i$ and $j$ respectively. These will be defined below. The non-negative function $g(\cdot,\cdot)$ represents the detailed mechanics of the competition. A number of specific forms are commonly used \cite{bladon}, but our method applies for a general choice. 

\begin{figure}[t]
%\vspace{3em}
\centerline{\includegraphics[width=0.52\textwidth]{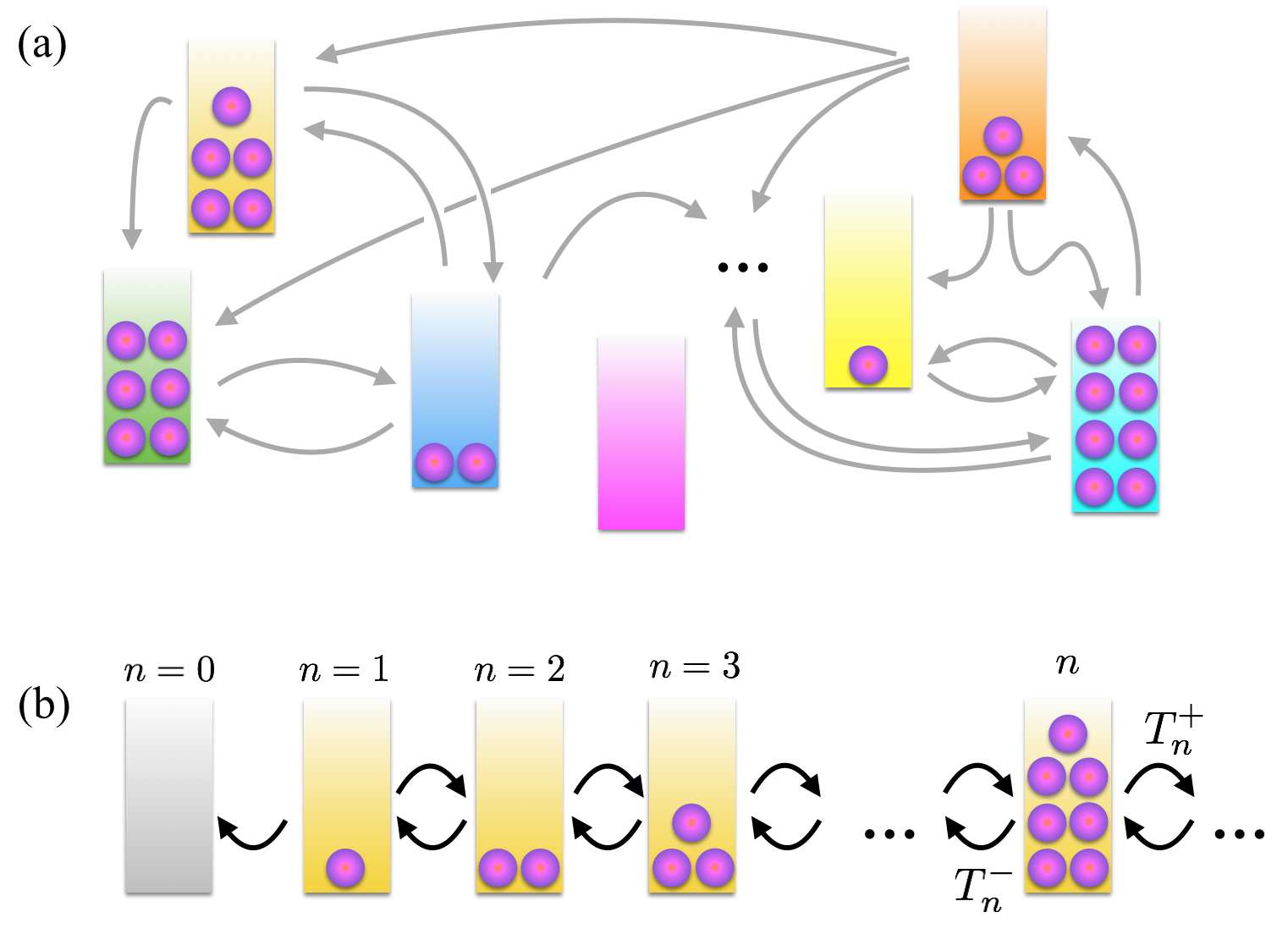}}
\vspace{0.5em}
\caption{(Colour on-line) (a) Original birth-death process. Each urn represents one species; birth-death events replace an individual of one species by an individual of another species.  We assume all-to-all interaction between species; not all arrows are drawn. Birth-death events between two species can occur provided that neither the originating urn (death) nor the the destination urn (reproduction) are empty. The species in the lower centre has reached extinction.  (b) The outcome of the path-integral analysis is an `effective' birth-death process on the domain $n=0,1,2,\dots$ for a single representative species. The state $n=0$ is absorbing. The effective birth and death rates, $T_n^\pm$, vary in time, and they depend on the history of the effective process.}
\label{fig:fig1}
\end{figure}

The model is illustrated in Fig. \ref{fig:fig1}(a). Each of the urns represents one species, and the figure shows the number of particles of each type. We assume all-to-all interaction so that birth-death events involving any pair of species are possible in principle; not all arrows are drawn in the figure. The reaction rates (\ref{eq:rates}) are such that species cannot be reintroduced once they have gone extinct; the urn in the lower centre of the figure illustrates this, there is no arrow pointing to it.
The fitnesses $\{f_i\}$ are set through pairwise interaction between species, $f_i=\Omega^{-1}\sum_j a_{ij}n_j$, with an interaction matrix ${\mathbb A}=(a_{ij})$. In evolutionary game theory this represents a two-player game with payoff matrix $\mathbb{A}$ \cite{multiverse}. The quenched uncertainty of the problem is contained in these interaction coefficients. Specifically, the $a_{ij}$ are Gaussian random variables with mean zero, and drawn before the population dynamics starts. They then remain fixed. We write $\overline{\cdots}$ for averages over the ensemble of matrices $\mathbb{A}$. The variance of the interaction coefficients is chosen as $\overline{a_{ij}^2}=1/S$; see also \cite{coolen, sollich, mpv, diederich2,gallajpa, gallafarmer}. A model parameter $\Gamma\in [-1,1]$ controls correlations between $a_{ij}$ and $a_{ji}$,  
\be
\overline{a_{ij}a_{ji}}=\frac{\Gamma}{S}.
\ee
Thus, $a_{ij}=a_{ji}$ with probability one for $\Gamma=1$, i.e. species cooperate. For $\Gamma=0$ interactions are uncorrelated, and for $\Gamma=-1$ we have a zero sum game, $a_{ij}=-a_{ji}$, with probability one; see also \cite{diederich2, gallajpa,gallafarmer}.
\\

\begin{figure}[t]
%\vspace{3em}
\centerline{\includegraphics[width=0.45\textwidth]{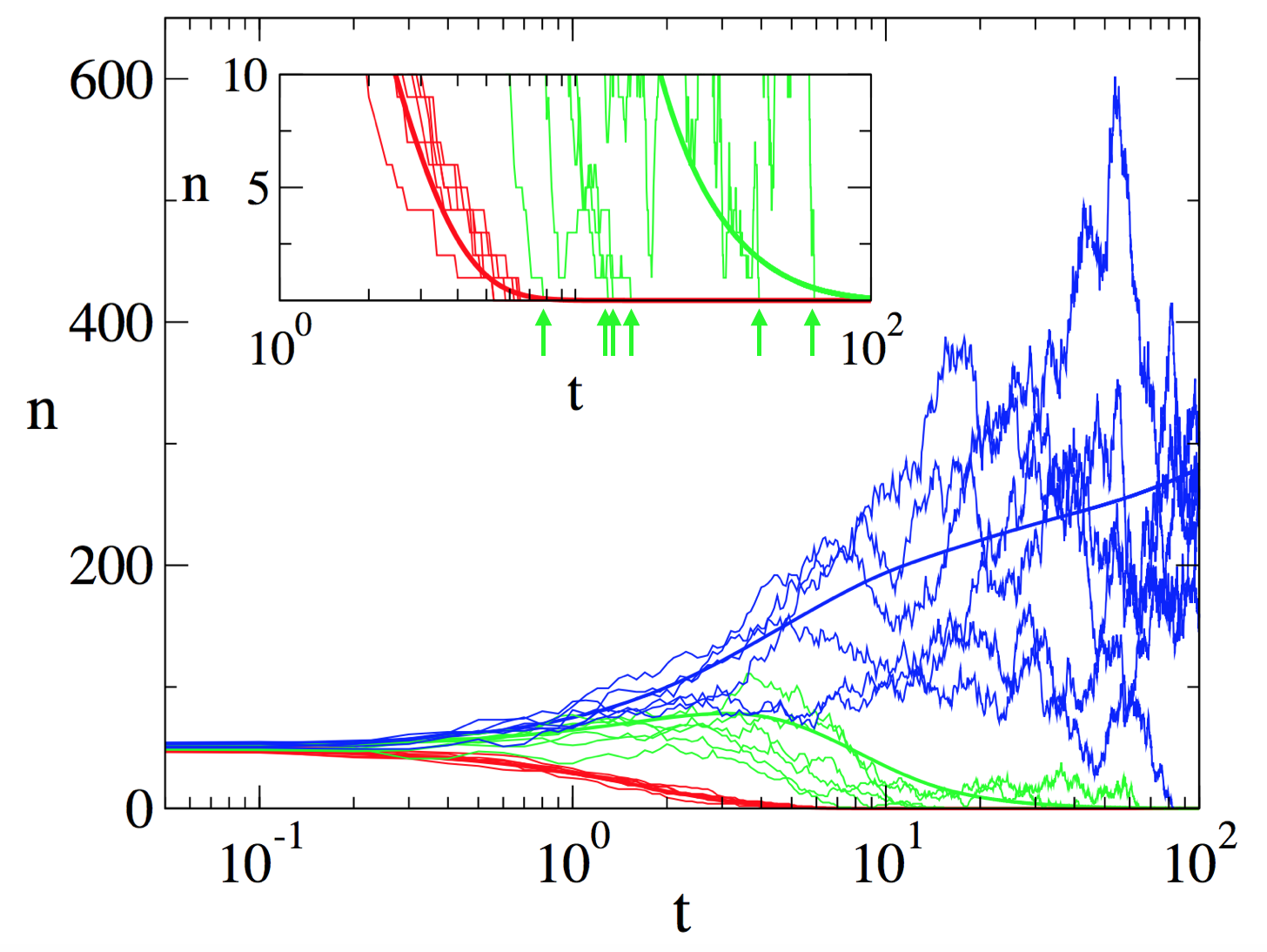}}
\vspace{0.5em}
\caption{(Colour on-line) Sample trajectories of the system with $S=100$. Noisy lines show several realisations of the stochastic dynamics ($\Omega=50$), for one realisation of the interaction coefficients $a_{ij}$, and for three randomly selected species. Smooth lines show the deterministic dynamics (\ref{eq:det}) \cite{remark}. Inset: Close-up near extinction. Times at which the species shown in green (light grey) goes extinct in the stochastic system are marked by arrows ($\Gamma=0.5$, $g(f_j,f_i)=\frac{1}{2}\left[1+\tanh(\beta f_j)\right]$, $\beta=1$).}
\label{fig:fig2}
\end{figure}

{\em Deterministic limit.} We keep $S$ finite for the time being. Writing $x_i=n_i/\Omega$ and focusing on a fixed sample of the $\{a_{ij}\}$ the dynamics becomes deterministic in the limit $\Omega\to\infty$. One then has
\be\label{eq:det}
\dot x_i=\sum_{j\neq i} T^\infty_{j\to i}(\bx)-T^\infty_{i\to j}(\bx).
\ee
The $T^\infty_{i\to j}(\bx)$ are obtained from the $T_{i\to j}(\bn)$ as in \cite{bladon}. Sample paths for one realisation of the disorder are shown in Fig. \ref{fig:fig2} for finite $\Omega$, and in the deterministic limit respectively. As seen in the inset a species may reach extinction at finite times in the stochastic system. Under the deterministic dynamics any $x_i$ can approach zero only asymptotically.
\\

{\em Objective of the analysis.} Our aim is to study the typical birth-death dynamics for a representative species after the average over the $\{a_{ij}\}$ has been carried out. To illustrate this it is helpful to focus on one species in the all-to-all geometry of Fig. \ref{fig:fig1}(a). Imagine now an average over the entire ensemble of possible matrices $\mathbb{A}$ is carried out. We ask what the process $n(t)$ for the focal species will typically look like. Naturally, it will be a birth-death dynamics on the space $n=0, 1, 2, \dots$, with an absorbing state at $n=0$; see Fig. \ref{fig:fig1}(b). It is the `effective' birth and death rates $T_n^\pm$ post disorder average which we wish to determine.

{\em Path-integral analysis and representative process.} Following  \cite{brettgalla, tauleaping} we discretise time,
\be
n_i(t+\Delta)=n_i(t)+\sum_j \left[k_{ji}(t)-k_{ij}(t)\right],
\ee
where the $\{k_{ij}(t)\}$ are Poissonian random variables with parameters $\lambda_{ij}(t)=\frac{\Delta}{\Omega S} n_i(t) n_j(t)g(f_j,f_i)$. We write $P(\bk)$ for their distribution \cite{note}. Continuous time is eventually restored by taking the limit $\Delta\to 0$ at the end of the calculation. We use a path-integral approach to proceed \cite{coolen, sollich, msr}. For a fixed sample of $\mathbb{A}$ the generating functional of the dynamics is given by
\BE\label{eq:genf}
Z[\boldpsi]&=&\!\!\int D\bn \prod_i p_0(n_i)\sum_{\bk} P(\bk) ~e^{i\Delta\sum_{it} \psi_i(t)n_i(t)}
 \\
&&\hspace{-2em}\times \prod_{i,t} \delta\left(n_i(t+\Delta)-n_i(t)-\sum_j \left[k_{ji}(t)-k_{ij}(t)\right]\right).\nonumber
\EE
The notation $\int D\bn$ represents the sum over all paths $\{\bn(t); t\geq t_0\}$. The distribution $p_0(n_i)$ is the initial condition for the $n_i$ at time $t_0$. They are assumed to be independent and identically distributed, with $\sum_i n_i=\Omega S$. In the thermodynamic limit $S\to \infty$, but keeping $\Omega=N/S$ finite, the generating functional can be averaged over the Gaussian disorder $\mathbb{A}$. The calculation is very technical, we report it in the Appendix. Details of the method can also be found in \cite{msr, coolen,sollich}. The final outcome of the generating functional analysis is a non-Markovian stochastic process, $n(t)$, for a representative species, and an associated representative fitness $f(t)$. The jump process $n\to n\pm 1$ is governed by rates
\BE
T^+&=&n\,b[f(t)],\nonumber \\
T^-&=&n\,d[f(t)],\label{eq:effrates}
\EE
where the per capita birth and death rates $b$ and $d$ are functions of the fitness $f(t)$, see below. The fitness in turn is of the form
\be\label{eq:fitnessprocess}
f(t)=\frac{\Gamma}{\Omega}\int_{t_0}^t dt' G(t,t')n(t')+\eta(t),
\ee
if the dynamics is started at $t_0$. The term $\eta(t)$ is coloured Gaussian noise with correlations to be described shortly along with the definition of the kernel $G$. The birth and death rates are given by
\BE
b(\phi)&=&\Omega^{-1}\avg{ng(\phi,f)}_*,\nonumber\\
d(\phi)&=&\Omega^{-1}\avg{ng(f,\phi)}_*.\label{eq:effratespercapita}
\EE
We have written $\avg{\dots}_*$ for averages over realisations of effective process, i.e., over the combined set $\{n, f,\eta\}$. It is important to note that no average over $\phi$ is performed in (\ref{eq:effratespercapita}). For the further analysis it is useful to describe the relation of the effective process to the original microscopic model. The generating functional calculation or dynamical mean field theory discards correlations between different species. However, it preserves the statistics of observables for single species. The average $\avg{\cdots}_\star$ over samples of the effective process is equivalent to the combined average over species {\em and} samples of the disorder in the microscopic model \cite{coolen, sollich}. This helps to interpret the expressions (\ref{eq:effratespercapita}), recalling that $n_jg(f_i,f_j)/N$ is the per capita rate with which individuals of type $i$ replace individuals of type $j$ in the original model. On the level of the dynamical mean field theory the average $\Omega^{-1}\avg{ng(\phi,f)}_\star$ is the equivalent of $N^{-1}\sum_j \overline{n_j g(\phi,f_j)}$, i.e. broadly speaking it is the reproductive success of a species with fitness $\phi$ in the ensemble post disorder average. The expression $\Omega^{-1}\avg{ng(f,\phi)}_\star$ is the rate by which such a species is displaced by other individuals.
\begin{figure}[t!!!]
\vspace{-2em}
\centerline{\includegraphics[width=0.52\textwidth]{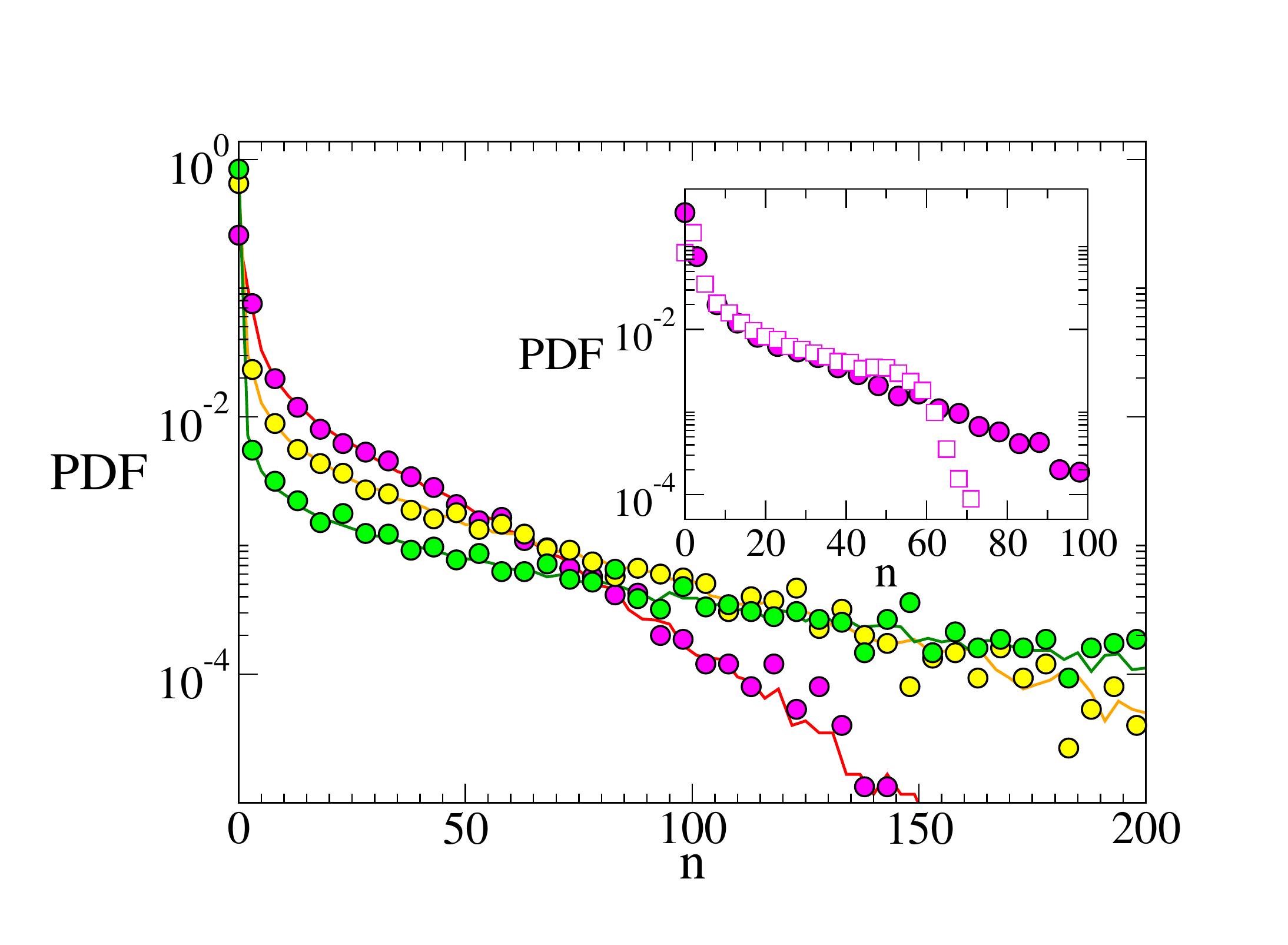}}
\vspace{0.5em}
\caption{(Colour on-line) Distribution of particle numbers, $n$, for a representative species ($\Gamma=-0.5, \beta=1$, $\Omega=10$). The distributions are shown at times $t=5$, $t=10$ and $t=20$ (from bottom to top at large $n$). Solid lines are from simulating the effective representative species dynamics ($2\times 10^5$ sample paths), markers from microscopic simulations ($S=300$, $50$ samples). Inset: Distribution of $n$ at $t=5$. Filled symbols are from simulation of the microscopic process, open markers from numerical integration of the deterministic rate equations ($100$ samples of the $\{a_{ij}\}$). }
\label{fig:fig3}
\end{figure}

\begin{figure}[t!!!]
\vspace{-1em}
\centerline{\includegraphics[width=0.55\textwidth]{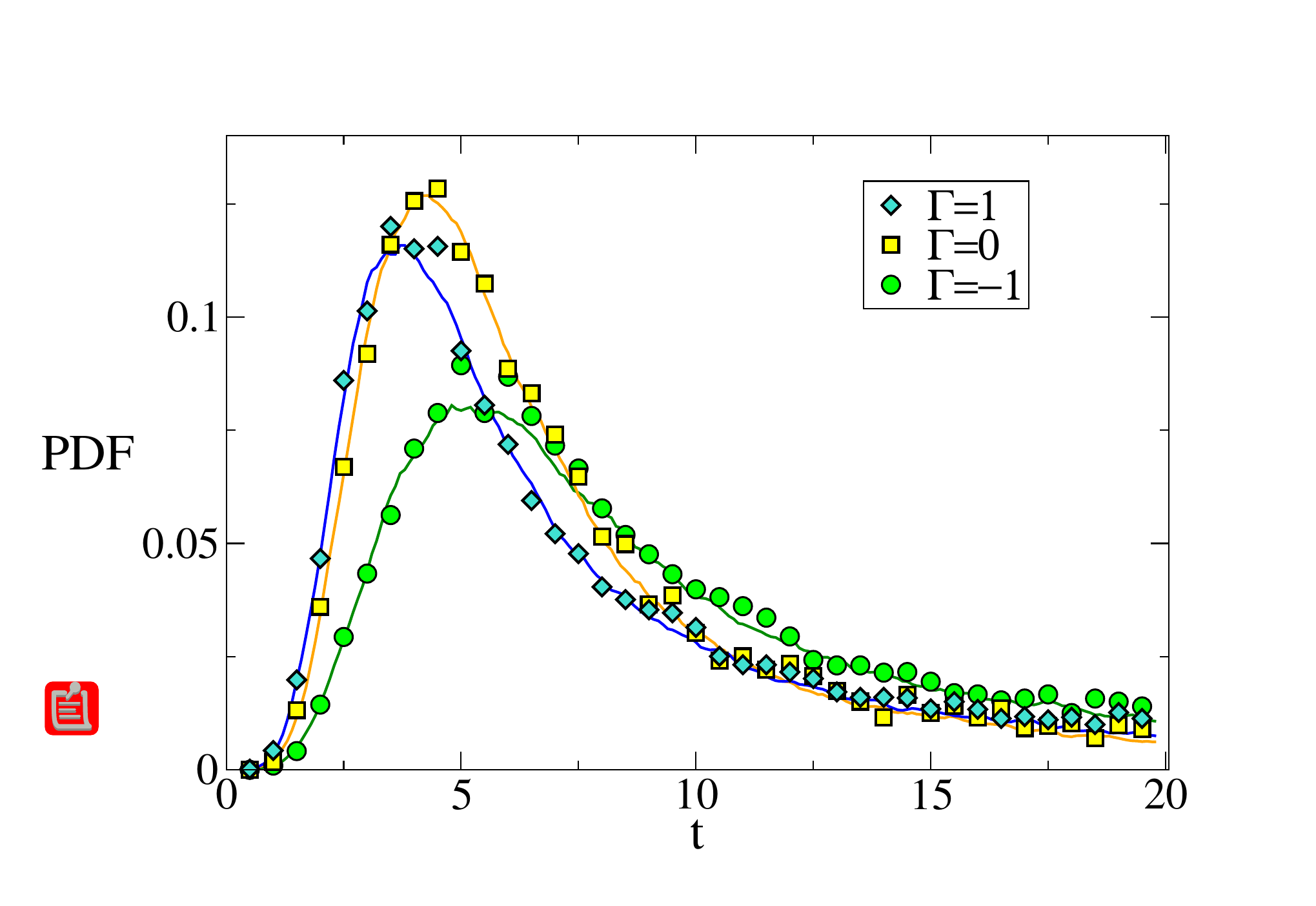}}
\vspace{0.5em}
\caption{(Colour on-line) Extinction time distribution for different choices of the correlation parameter $\Gamma$ ($\beta=1$, $\Omega=10$). Solid lines show results from simulating the effective representative species dynamics ($2\times 10^5$ sample paths), markers are from microscopic simulations ($S=300$, averaged over $50$ samples).}
\label{fig:fig4}
\end{figure}

The kernel $G(t,t')$ in (\ref{eq:fitnessprocess}) is a response function, to be determined from
\be\label{eq:g}
G(t,t')=\Omega^{-1}\avg{\frac{\delta n(t)}{\delta \eta(t')}}_*.
\ee
The correlator $C(t,t')\equiv \avg{\eta(t)\eta(t')}_*$ finally obeys the relation
\be\label{eq:c}
C(t,t')=\Omega^{-2}\avg{n(t)n(t')}_*,
\ee
see the Appendix for details. The effective process (\ref{eq:effrates})-(\ref{eq:effratespercapita}) together with the self-consistenty relations (\ref{eq:g},\ref{eq:c}) determine the macroscopic order parameters $C$ and $G$. \\
{\em Self-consistent simulation method.} A simulation method for the effective birth-death process can be devised combining ideas of the Gillespie algorithm \cite{Gillespie1, Gillespie2} with the technique proposed by Eissfeller and Opper to simulate the dynamical mean field theory of spin glasses \cite{eissopp1,eissopp2}. This technique iteratively generates sample paths of the effective process. From these the macroscopic order parameters are obtained self-consistently. After initialisation the simulation broadly proceeds along the following steps: (i) Assume sample paths and macroscopic order parameters $C, G$ have been generated up to time $t$. Create Gaussian noise $\eta$ with appropriate correlations, and use it to generate the fitness $f(t)$ via Eq. (\ref{eq:fitnessprocess}) for each sample; (ii) Again for each sample use Eqs. (\ref{eq:effrates},\ref{eq:effratespercapita}) to obtain transition rates for the birth-death process. Use these to advance the sample paths, $n(t)$, with the time-leaping Gillespie method; (iii) Use (\ref{eq:g},\ref{eq:c}) to update the dynamical order parameters   as averages over paths; (iv) Iterate. The calculation of the response function $G$, and the generation of the coloured Gaussian noise $\eta(t)$ require intermediate steps \cite{eissopp1, eissopp2}. These are are described in the Appendix, along with further details of the algorithm.
\\
{\em Test against simulations.} To demonstrate the simulation method we focus on the sinusoidal birth rate $g(f_j,f_i)=\frac{1}{2}\left[1+\tanh(\beta f_j)\right]$ in Figs. \ref{fig:fig3} and \ref{fig:fig4}.  The parameter $\beta\geq 0$ is a selection strength \cite{bladon,traulsenfermi}. This functional form is particularly simple as it only involves the fitness of the reproducing species; a randomly chosen individual is removed at each birth event. We have tested other choices, in particular $g(f_i,f_j)= 1/[1+\exp(-2\beta(f_i-f_j))]$, leading to what is sometimes referred to as the `Fermi' process \cite{traulsenfermi, bladon}. In Fig. \ref{fig:fig3} we show the distribution of particle numbers, $P_t(n)$, at different times of the evolutionary process. This quantity is the probability to find precisely $n$ individuals of a randomly chosen species in the population at time $t$. Equivalently it is the probability for a sample path of the effective dynamics to be in state $n$ at the $t$. The data in Fig. \ref{fig:fig3} shows good agreement between simulations of the effective process and those of the original microscopic model. As seen in the figure the distribution $P_t(n)$ broadens with time, as some species go extinct, while others are present in the resulting `condensate'  with relatively large particle numbers \cite{knebel}. In the inset of Fig. \ref{fig:fig3} we compare the outcome of the individual-based model with that of the deterministic limit (\ref{eq:det}).  The distribution of the stochastic model is broader and with a non-zero extinction probability at finite times \cite{remark1}. We show extinction time distributions of the stochastic model in Fig. \ref{fig:fig4} for different choices of the competition parameter $\Gamma$. The data confirms again the validity of the simulation method for the effective process. A more detailed discussion of fixation in random games and the possible biological implications is not the objective of the present work, and will be presented elsewhere. 

{\em Strengths and limitations.}
The strength of the algorithm is the ability to simulate the typical dynamics of a representative species, rather than to carry out simulations sample by sample of the quenched disorder. The technique is free from finite size effects, as the effective dynamics is derived in the limit of infinitely many species, $S\to\infty$. Remaining sampling errors are statistical, but not systematic \cite{eissopp1,eissopp2}. The method requires a discretisation of time, mainly for the necessary linear algebra. Further drawbacks include the restriction to Gaussian couplings, and the computational cost in generating the response matrix and the coloured noise. Simulations are only practical for a few hundred time steps; for our evolutionary system we use time leap of $\Delta=0.1$ generations, so we have only explored the method for times up to tens of generations.

{\em Diffusion approximation.} The transition rates (\ref{eq:effrates}) can be used as a starting point to derive a diffusion approximation \cite{vankampen, gardiner} for the representative birth-death process. This process is defined on the domain $n=0,1,2,\dots $, and the scale for the typical number of individuals is set by $\Omega$, viz. $\avg{n(t)}_\star=\Omega$. The diffusion approximation formally consists of an expansion in powers of $\Omega^{-1}$ at the level of the generating functional {\em post} disorder average; see the Appendix for details. Expanding to sub-leading order one finds
\be
\dot x = {\cal T}^+(x)-{\cal T}^-(x)+\sqrt{\frac{{\cal T}^+(x)+{\cal T}^-(x)}{\Omega}}\xi,
\ee
where where $\avg{\xi(t)\xi(t')}=\delta(t-t')$. The rates ${\cal T}^\pm(x)=\Omega^{-1}T^\pm (n=x\Omega)$ are obtained from Eqs. (\ref{eq:effrates}, \ref{eq:fitnessprocess}). They depend on the fitness defined in Eq. (\ref{eq:fitnessprocess}). For the Fermi process and assuming weak selection ($\beta\ll 1$) we have $g(f_i,f_j)=1/[1+\exp\{-2\beta(f_i-f_j)\}]\approx [1+\beta(f_i-f_j)]/2$. This leads to
\be
\dot x = \beta x [f-\avg{x f}_\star]+\sqrt{\frac{x}{\Omega}}\xi, 
\ee
 The coloured noise $\eta(t)$ in the definition of the fitness [Eq. (\ref{eq:fitnessprocess})] originates from the quenched disorder. The Gaussian white noise $\xi(t)$ and the prefactor proportional to $\Omega^{-1/2}$ reflects the intrinsic noise of the problem in the diffusion approximation.  The noise term containing $\xi(t)$ is multiplicative, and $x=0$ an absorbing state. In the limit $\Omega\to\infty$ one finds the effective process, $\dot x = \beta x [f-\avg{xf}_\star]$ of the random replicator equations, previously studied in \cite{diederich, diederich2, gallajpa}.
\\

{\em Summary and discussion.} Many biological processes involve an intricate interplay of intrinsic noise and quenched uncertainty. We have used a generating functional approach to derive an effective birth-death process, valid after integrating out the quenched disorder. It describes the typical jump process for a representative species. The disorder leads to history dependent transition rates and coloured noise. In the diffusion approximation we obtain a non-Markovian stochastic differential equation; it reduces to the known effective dynamics for the random replicator equations in the limit of vanishing intrinsic noise. We have combined the Gillespie algorithm with the method by Eissfeller and Opper to propose a simulation technique for the effective birth-death process post disorder average. This numerical method has limitations and at present it only applies for all-to-all interactions. However, it allows one to simulate the `typical' birth-death process for a representative species, rather than to study the disorder sample by sample. This can be a useful tool to understand how quenched uncertainty propagates and interacts with other types of noise. We also believe that the effective jump process provides an interesting mathematical view on reaction systems with intrinsic noise and random rate constants.  Natural next steps include extensions to random reaction networks, and to models in which not only the rate constants are uncertain, but also the effects of the reactions.
%%%%%%%%%%%%%%%%%%%%%%%%%%%%%%%%%%%%%%%%%%%%%%%%%%%%%%%%%%%%%%%%%%

\begin{acknowledgments} 
Support by the EPSRC is gratefully acknowledged (grant reference EP/K037145/1). I would like to thank Johannes Knebel for discussions, and the Group of Nonlinear Physics, University of Santiago de Compostela, Spain for hospitality. \end{acknowledgments}

\widetext
\begin{appendix}

\section{Path-integral analysis}
 \subsection{Model definitions}
We consider a population of $N=S\times\Omega$ individuals, each of which can be of one of $S$ species ($i=1,\dots,S$). We write $n_i$ for the number of individuals of type $i$, and we consider the following dynamics:
\be
T_{i\to j}=\frac{n_i n_j}{N}g(f_j,f_i),
\ee
where $g(\cdot, \cdot)$ is an arbitrary non-negative function. The fitness variables $f_i$ are defined via $f_i=\sum_j a_{ij}\frac{n_j}{\Omega}$, and the $a_{ij}$ are Gaussian quenched random variables, with
\be
\overline{a_{ij}}=0,~~~\overline{a_{ij}^2}=\frac{1}{S}, ~~~ \overline{a_{ij}a_{ji}}=\frac{\Gamma}{S}.
\ee
As a first step we discretise time into time steps $\Delta$, and introduce reduced variables $x_i=\frac{n_i}{\Omega}$. We assume that reaction rates are constant in each time interval $\Delta$, akin to a $\tau$-leaping procedure of the Gillespie algorithm \cite{tauleaping}. The number of reactions $i\to j$ in a given time interval is then a Poissonian random variable with rate $T_{i\to j}\Delta$. Discrete time steps are introduced for convenience, we will restore continuous time in due course by taking the limit $\Delta\to 0$. In discrete time the dynamics can be written as 
\be
x_i(t+\Delta)=x_i(t)+\frac{1}{\Omega}\sum_j \left[k_{ji}(t)-k_{ij}(t)\right],
\ee
where the $\{k_{ij}(t)\}$ are Poissonian random variables with parameters
\be
\lambda_{ij}(t)=\Delta\frac{\Omega}{S} x_i(t) x_j(t)g[f_j(\bn(t)),f_i(\bn(t))].
\ee

\subsection{Generating functional}
The generating functional reads (prior to carrying out the disorder average)
\BE
Z[\boldpsi]&=&\int Dx D\widehat x \sum_{\bf k} P(\bk) \left[\prod_{i} p_0(x_i)\right]\exp\left(i\Delta\sum_{i,t} x_i(t)\psi_i(t)\right)\times \exp\left(i\sum_{i,t}\widehat x_i(t)\left[x_i(t+\Delta)-x_i(t)\right]\right)\nonumber \\
&&\times \exp\left(-i\frac{1}{\Omega}\sum_{i,t}\widehat x_i(t)\sum_j \left[k_{ji}(t)-k_{ij}(t)\right]\right).
\EE
The initial condition for the dynamics $p(\bn)$ at time $t=t_0$ is assumed to factorise, with all $x_i$ identically distributed with density $p_0(x_i)$ with unit mean reflecting the condition $\sum_i n_i =\Omega S$. The integral (sum) over the variables $\bx$ and the Poissonian random variables $\bk$ is to be carried out iteratively, see also \cite{brettgalla}. The term containing the dynamic noise (i.e., the variables $k_{ij}$)  can be written as
\BE
\exp\left(-i\frac{1}{\Omega}\sum_{i,t}\widehat x_i(t)\sum_j \left[k_{ji}(t)-k_{ij}(t)\right]\right)=\prod_{ij,t} \exp\left(-ik_{ij}(t)\frac{1}{\Omega}\left[\widehat x_j(t)-\widehat x_i(t)\right]\right).
\EE
We now proceed along the lines of \cite{brettgalla} and perform the average over the $k_{ij}(t)$. We use the identity
\BE
\sum_{k=0}^\infty e^{-\lambda}\frac{\lambda^k}{k!}e^{-ikx}=e^{-\lambda+\lambda e^{-ix}}.
\EE
This leads to
\BE
Z[\boldpsi]&=&\int Dx D\widehat x \left[\prod_{i} p_0(x_i)\right]\exp\left(i\Delta\sum_{i,t} x_i(t)\psi_i(t)\right)\ \exp\left(i\sum_{i,t}\widehat x_i(t)\left[x_i(t+\Delta)-x_i(t)\right]\right)\nonumber \\
&&\times \exp\left(\sum_{ij,t}\left(-\lambda_{ij}(t)+\lambda_{ij}(t)e^{-\frac{i}{\Omega}\left[\widehat x_j(t)-\widehat x_i(t)\right]}\right)\right),
\EE
which can be written as
\BE
Z[\boldpsi] &=&\int Dx D\widehat x   \left[\prod_{i} p_0(x_i)\right] \exp\left(i\Delta\sum_{i,t} x_i(t)\psi_i(t)\right) \exp\left(i\sum_{i,t}\widehat x_i(t)\left[x_i(t+\Delta)-x_i(t)\right]\right)\nonumber \\
&&\times \exp\left(-\Delta\sum_{ij,t} \frac{\Omega}{S} x_i x_j g(f_j,f_i)\left(1-e^{-\frac{i}{\Omega}\left[\widehat x_j(t)-\widehat x_i(t)\right]}\right)\right) 
\EE
The fitness variables are mere placeholders in this expression. Introducing $f_j(t)=\sum_k a_{jk}x_k(t)$ via appropriate delta-functions we can write this as
\BE
Z[\boldpsi]&=&\int Dx D\widehat x Df D\widehat f  \left[\prod_{i} p_0(x_i)\right]\exp\left(i\Delta\sum_{i,t} x_i(t)\psi_i(t)\right)\ \exp\left(i\sum_{i,t}\widehat x_i(t)\left[x_i(t+\Delta)-x_i(t)\right]\right)\nonumber \\
&& \times \exp\left(-\Delta\sum_{ij,t}\frac{\Omega}{S} x_i x_jg(f_j,f_i)\left(1-e^{-\frac{i}{\Omega}\left[\widehat x_j(t)-\widehat x_i(t)\right]}\right)\right) \nonumber \\
&& \times \exp\left(i\Delta\sum_{i,t} \widehat f_i(t)f_i(t)-i\Delta\sum_{i,t}\widehat f_i(t) \sum_j a_{ij} x_j(t)\right),
\EE
where we suppress the time dependence of the arguments of $g$.
\subsection{Disorder average}
We next carry out the average over the Gaussian quenched disorder (i.e., over the $\{a_{ij}\})$. Their statistics are indicated above. This average is denoted by an overbar $\overline{\cdots}$. We have
\BE
&&\overline{\exp\left(-i\Delta\sum_{i<j} \sum_t\left[a_{ij}\widehat f_i(t) x_j(t)+a_{ji}\widehat f_j(t) x_i(t)\right]\right)}\nonumber \\
&=&\exp\left(-\frac{1}{2S}\Delta^2\sum_{ij}\sum_{t,t'} \left[\widehat f_i(t)\widehat f_i(t') x_j(t)x_j(t')+\Gamma \widehat f_i(t)x_i(t')\widehat f_j(t')x_j(t)\right]\right).
\EE
Using this we find
\BE
\overline{Z[\boldpsi]}&=&\int Dx D\widehat x Df D\widehat f  \left[\prod_{i} p_0(x_i)\right] \exp\left(i\Delta\sum_{i,t} x_i(t)\psi_i(t)\right)\ \exp\left(i\sum_{i,t}\widehat x_i(t)\left[x_i(t+\Delta)-x_i(t)\right]\right)\nonumber \\
&&\times \exp\left(-\Delta\sum_{ij,t}\frac{\Omega}{S} x_i x_jg(f_j,f_i)\left(1-e^{-\frac{i}{\Omega}\left[\widehat x_j(t)-\widehat x_i(t)\right]}\right)\right)  \times \exp\left(i\Delta\sum_{i,t} \widehat f_i(t)f_i(t)\right)\nonumber \\
&&\times \exp\left(-\frac{1}{2S}\Delta^2\sum_{ij}\sum_{t,t'} \left[\widehat f_i(t)\widehat f_i(t') x_j(t)x_j(t')+\Gamma \widehat f_i(t)x_i(t')\widehat f_j(t')x_j(t)\right]\right).
\EE
We proceed by introducing macroscopic order parameters as follows
\BE
L(t,t')&=&\frac{1}{S}\sum_i \widehat f_i(t) \widehat f_i(t'), \nonumber \\
K(t,t')&=&\frac{1}{S}\sum_i x_i(t)\widehat f_i(t'), \nonumber \\
C(t,t')&=&\frac{1}{S}\sum_i x_i(t) x_i(t').
\EE
This results in 
\BE
\overline{Z[\boldpsi]}&=&\int Dx D\widehat x Df d\widehat f DC D\widehat C DL D\widehat L D K D\widehat K  \left[\prod_{i} p_0(x_i)\right]\exp\left(i\Delta\sum_{i,t} x_i(t)\psi_i(t)\right) \nonumber \\
&&\times\exp\left(i\sum_{i,t}\widehat x_i(t)\left[x_i(t+\Delta)-x_i(t)\right]\right)\nonumber \\
&&\times \exp\left(-\Delta\sum_{ij,t}\frac{\Omega}{S} x_i x_j g(f_j,f_i)\left(1-e^{-\frac{i}{\Omega}\left[\widehat x_j(t)-\widehat x_i(t)\right]}\right)\right)  \times \exp\left(i\Delta\sum_{i,t} \widehat f_i(t)f_i(t)\right)\nonumber \\
&&\times \exp\left(-\frac{1}{2}S\Delta^2\sum_{t,t'} \left[L(t,t')C(t,t')+\Gamma K(t,t') K(t',t)\right]\right)\nonumber \\
&&\times \exp\left(iS\Delta^2 \sum_{t,t'}\left[\widehat C(t,t') C(t,t')+\widehat K(t,t') K(t,t')+\widehat L(t,t') L(t,t')\right]\right) \nonumber \\
&&\times \exp\left(-i\Delta^2\sum_{i,t,t} \left[\widehat C(t,t')x_i(t)x_i(t')+\widehat K(t,t')x_i(t)\widehat f_i(t')+\widehat L(t,t')\widehat f_i(t)\widehat f_i(t')\right]\right).
\EE

Next, we look at the term
\BE
&&\exp\left(-\Delta\sum_{ij,t}\frac{\Omega}{S} x_i x_j \left\{g(f_j,f_i)\left(1-e^{-\frac{i}{\Omega}\left[\widehat x_j(t)-\widehat x_i(t)\right]}\right)\right\}\right) \nonumber \\
&=&\exp\left(\Delta\sum_{ij,t}\frac{\Omega}{S} x_i x_j \left\{g(f_j,f_i)\sum_{\ell=1}^\infty \frac{(-i)^\ell}{\ell !}\frac{(\widehat x_j(t)-\widehat x_i(t))^\ell}{\Omega^\ell}\right\}\right)\nonumber \\
&=&\exp\left(\Delta\sum_{\ell=1}^\infty \frac{(-1)^\ell}{\ell ! \Omega^{\ell-1}}\sum_{m=0}^\ell \left(\begin{array}{c} \ell \\ m \end{array}\right) \frac{1}{S}\sum_{ij,t} x_i x_j \left\{g(f_j,f_i)(i\widehat x_j(t))^m (-i\widehat x_i(t))^{\ell-m}\right\}\right)
.
\EE
\\
We now examine the term $\frac{1}{S}\sum_{ij,t} x_i x_j \left\{g(f_j,f_i)(i\widehat x_j(t))^m (-i\widehat x_i(t))^{\ell-m}\right\}$ in more detail. The function $g(f_j,f_i)$ can be written as a power series $g(f_j,f_i)=\sum_{\mu\nu} g_{\mu\nu}f_j^\mu f_i^\nu$, with suitable coefficients, $g_{\mu\nu}=g_{\mu\nu}(\beta)$. To keep the notation under control at least to some degree, we suppress the dependence on $\beta$.
\\

Using this series expansion, it is sufficient to analyse terms of the form $\frac{1}{S}\sum_{ij} x_i x_j f_j^\mu f_i^\nu (i\widehat x_j(t))^m (-i\widehat x_i(t))^{\ell-m}$. In the thermodynamic limit ($S\to\infty$) such terms will only contribute to the expression in the exponential when $m=0$ or $m=\ell$. To see this consider a case in which $0<m<\ell$. We then have
\BE
&&\frac{1}{S}\sum_{ij} x_i(t) x_j(t) f_j(t)^\mu f_i(t)^\nu (i\widehat x_j(t))^m (-i \widehat x_i(t))^{\ell-m}\nonumber \\
&=&\sum_i (-i\widehat x_i(t))^{\ell-m} x_i(t) f_i(t)^\nu \left[\frac{1}{S}\sum_j x_j(t) f_j(t)^\mu (i\widehat x_j(t))^m\right].
\EE

In the limit $S\to \infty$ and carrying out a saddle point integration, the term in the square bracket is an equal-time response function for $m>0$, and zero by causality \cite{coolen, sollich, eissopp1, eissopp2, diederich2} Alternatively, we can write
\BE
&&\frac{1}{S}\sum_{ij} x_i(t) x_j(t) f_j(t)^\mu f_i(t)^\nu (i\widehat x_j(t))^m (-i\widehat x_i(t))^{\ell-m}\nonumber \\
&=&\sum_j x_j(t) f_j(t)^\mu (i\widehat x_j(t))^m \left[\frac{1}{S}\sum_i (-i\widehat x_i(t))^{\ell-m} x_i(t) f_i(t)^\nu\right],
\EE
and the expression in the square brackets is recognised as an equal-time response function again for $m<\ell$, and vanishes at the saddle point.

Thus we are left with the terms $m=0$ and $m=\ell$, and so
\BE
&&\exp\left(-\Delta\sum_{\mu\nu} g_{\mu\nu} \sum_{ij,t}\frac{\Omega}{S} x_i(t) x_j(t) \left\{f_j(t)^\mu f_i(t)^\nu\left(1-e^{-\frac{i}{\Omega}\left[\widehat x_j(t)-\widehat x_i(t)\right]}\right)\right\}\right) \nonumber \\
&=&\exp\left(\sum_{\mu\nu} g_{\mu\nu}\sum_{\ell=1}^\infty \frac{(-1)^\ell}{\ell ! \Omega^{\ell-1}}\left\{ \frac{1}{S}\Delta\sum_{ij,t} x_i(t) x_j(t) \left\{ f_j(t)^\mu f_i(t)^\nu (-i\widehat x_i(t))^{\ell}\right\}\right.\right.\nonumber \\
&&\hspace{12em}\left.\left.+\frac{1}{S}\Delta\sum_{ij,t} x_i(t) x_j(t) \left\{f_j(t)^\mu f_i(t)^\nu (i\widehat x_j(t))^\ell \right\}\right\}\right)\nonumber \\
&=&\exp\left(\sum_{\mu\nu} g_{\mu\nu}\sum_{\ell=1}^\infty \frac{1}{\ell ! \Omega^{\ell-1}}\left\{ \frac{1}{S}\Delta\sum_{ij,t} x_i(t) x_j(t) \left\{f_j(t)^\mu f_i(t)^\nu (i\widehat x_i(t))^{\ell}\right\}\right.\right.\nonumber \\
&&\hspace{12em} \left.\left.+\frac{1}{S}\Delta\sum_{ij,t} x_i(t) x_j(t) \left\{f_j(t)^\nu f_i(t)^\mu(-i\widehat x_i(t))^\ell \right\}\right\}\right),  \label{eq:help}
\EE
where we have relabelled indices ($i\leftrightarrow j$) in the last term. 

We next introduce 
\be
 R_\mu(t)= \frac{1}{S}\sum_j x_j(t) f_j(t)^\mu.
 \ee
 These are shorthands for the time being, suitable delta-functions will be inserted in the generating functional below. The expression in Eq. (\ref{eq:help}) can then be written as
\BE
&& \exp\left(\sum_{\mu\nu}g_{\mu\nu}\Omega\sum_{\ell=1}^\infty \frac{1}{\ell ! \Omega^{\ell}}\left\{\Delta \sum_{i,t}  R_\mu(t) x_i(t)f_i(t)^\nu (i\widehat x_i(t))^{\ell}+\Delta\sum_{i,t} R_\nu(t) x_i(t)  f_i(t)^\mu  (-i\widehat x_i(t))^\ell \right\}\right)\nonumber \\
&=&\exp\left(\sum_{\mu\nu}g_{\mu\nu}\Omega\Delta\sum_{it}\left[R_\mu(t)x_i(t)f_i(t)^\nu\left\{e^{\frac{i\widehat x_i(t)}{\Omega}}-1\right\}+R_\nu(t)x_i(t) f_i(t)^\mu \left\{e^{-\frac{i\widehat x_i(t)}{\Omega}}-1\right\}\right]\right).
\EE
 So finally, we conclude
  \BE
&&\exp\left(-\Delta\sum_{ij,t}\frac{\Omega}{S} x_i x_j \left\{g(f_j,f_i)\left(1-e^{-\frac{i}{\Omega}\left[\widehat x_j(t)-\widehat x_i(t)\right]}\right)\right\}\right) \nonumber \\
&=&\exp\left(\sum_{\mu\nu}g_{\mu\nu}\Omega\Delta\sum_{it}\left[R_\mu(t)x_i(t)f_i(t)^\nu\left\{e^{\frac{i\widehat x_i(t)}{\Omega}}-1\right\}+R_\nu(t)x_i(t) f_i(t)^\mu \left\{e^{-\frac{i\widehat x_i(t)}{\Omega}}-1\right\}\right]\right).
\EE
 with the above expressions for the $\{R_\mu\}$. The full generating functional (post disorder average) then reads
 \BE
\overline{Z[\boldpsi=0]}&=&\int Dx D\widehat x Df d\widehat f DC D\widehat C DL D\widehat L D K D\widehat K DR D\widehat R  \left[\prod_{i} p_0(x_i)\right]\nonumber \\
&&\times\exp\left(i\sum_{i,t}\widehat x_i(t)\left[x_i(t+\Delta)-x_i(t)\right]\right)\nonumber \\
&&\times \exp\left(i\Delta\sum_{i,t} \widehat f_i(t)f_i(t)+iS\Delta\sum_t \sum_\mu \widehat R_\mu(t)R_\mu(t)\right)\nonumber \\
&&\times \exp\left(-i\Delta\sum_\mu \sum_t \widehat R_\mu (t)\sum_i x_i(t)f_i(t)^\mu\right)\nonumber \\
&&\times\exp\left(\sum_{\mu\nu}g_{\mu\nu}\Omega\Delta\sum_{it}\left[R_\mu(t)x_i(t)f_i(t)^\nu\left\{e^{\frac{i\widehat x_i(t)}{\Omega}}-1\right\}+R_\nu(t)x_i(t) f_i(t)^\mu \left\{e^{-\frac{i\widehat x_i(t)}{\Omega}}-1\right\}\right]\right)\nonumber\\
&&\times \exp\left(-\frac{1}{2}S\Delta^2\sum_{t,t'} \left[L(t,t')C(t,t')+\Gamma K(t,t') K(t',t)\right]\right)\nonumber \\
&&\times \exp\left(iS\Delta^2 \sum_{t,t'}\left[\widehat C(t,t') C(t,t')+\widehat K(t,t') K(t,t')+\widehat L(t,t') L(t,t')\right]\right) \nonumber \\
&&\times \exp\left(-i\Delta^2\sum_{i,t,t} \left[\widehat C(t,t')x_i(t)x_i(t')+\widehat K(t,t')x_i(t)\widehat f_i(t')+\widehat L(t,t')\widehat f_i(t)\widehat f_i(t')\right]\right).
\EE
We have set the source term to zero for convenience. This can be written as
\BE
\overline{Z}&=&\int  DC D\widehat C DL D\widehat L D K D\widehat K DR D\widehat R~ e^{S\left(\Phi+\Psi+\Upsilon\right)},
\EE
where
\BE
\Phi&=&-\frac{1}{2}\Delta^2\sum_{t,t'} \left[L(t,t')C(t,t')+\Gamma K(t,t') K(t',t)\right],\nonumber\\
\Psi&=&i\Delta^2\sum_{t,t'}\left[\widehat C(t,t') C(t,t')+\widehat K(t,t') K(t,t')+\widehat L(t,t') L(t,t')\right] +i\Delta\sum_t\sum_\mu \widehat R_\mu(t)R_\mu(t),
\EE
and with the single effective species measure
\BE
\Upsilon&=&\ln\bigg[ \int Dx D\widehat x Df D\widehat f ~p_0(x)\exp\left(i\sum_{t}\widehat x(t)\left[x(t+\Delta)-x(t)\right]\right) \exp\left(i\Delta\sum_{t} \widehat f(t)f(t)\right)\nonumber \\
&&\times \exp\left(-i\Delta\sum_t \sum_\mu\widehat R_\mu(t) x(t)f(t)^\mu\right)\nonumber \\
&&\times\exp\left(\sum_{\mu\nu}g_{\mu\nu}\Omega\Delta\sum_{t}\left[R_\mu(t)x(t)f(t)^\nu\left\{e^{\frac{i\widehat x(t)}{\Omega}}-1\right\}+R_\nu(t)x(t) f(t)^\mu \left\{e^{-\frac{i\widehat x(t)}{\Omega}}-1\right\}\right]\right)\nonumber \\
&& \times\exp\left(-i\Delta^2\sum_{t,t'} \left[\widehat C(t,t')x(t)x(t')+\widehat K(t,t')x(t)\widehat f(t')+\widehat L(t,t')\widehat f(t)\widehat f(t')\right]\right)\bigg].\label{eq:upsilon}
\EE
\subsection{Saddle-point integration}
Next we carry out the saddle-point integration in the limit $S\to \infty$ (at finite $\Omega=N/S$). We get
\BE
\frac{\delta}{\delta C(t,t')}\left[\Phi+\Psi+\Upsilon\right]= 0&\Rightarrow& i\widehat C(t,t')=\frac{1}{2}L(t,t'),\nonumber \\
\frac{\delta}{\delta L(t,t')}\left[\Phi+\Psi+\Upsilon\right]= 0&\Rightarrow& i\widehat L(t,t')=\frac{1}{2}C(t,t'),\nonumber \\
\frac{\delta}{\delta K(t,t')}\left[\Phi+\Psi+\Upsilon\right]= 0&\Rightarrow& i\widehat K(t,t')=\Gamma K(t',t).
\EE
Furthermore
\BE
\frac{\delta}{\delta \widehat C(t,t')}\left[\Phi+\Psi+\Upsilon\right]= 0 &\Rightarrow&C(t,t')=\avg{x(t)x(t')}_\Upsilon,\nonumber \\
\frac{\delta}{\delta \widehat L(t,t')}\left[\Phi+\Psi+\Upsilon\right]= 0 &\Rightarrow& L(t,t')=\avg{\widehat f(t)\widehat f(t')}_\Upsilon,\nonumber \\
\frac{\delta}{\delta \widehat K(t,t')}\left[\Phi+\Psi+\Upsilon\right]= 0 &\Rightarrow& K(t,t')=\avg{x(t)\widehat f(t')}_\Upsilon,
\EE
where $\avg{\cdots}_\Upsilon$ defines a representative-particle measure
\be
\avg{F(x,\widehat x, f, \widehat f)}_\Upsilon=\frac{ \int Dx D\widehat x Df D\widehat f ~p_0(x)\exp(\cdots) F(x,\widehat x, f, \widehat f)}{\int Dx D\widehat x Df D\widehat f ~p_0(x)\exp(\cdots)},
\ee
where the exponential is that inside the logarithm in Eq. (\ref{eq:upsilon}).

One finds that $L(t,t')=0$ as usual \cite{coolen, sollich}.
\\

Next we have
\BE
\frac{\delta}{\delta R_\mu(t)}\left[\Phi+\Psi+\Upsilon\right]= 0&\Rightarrow& i\widehat R_\mu(t)=-\Omega\sum_{\nu} g_{\mu\nu}\avg{x(t)f(t)^\nu\left[e^{\frac{i\widehat x(t)}{\Omega}}-1\right]}_\Upsilon\nonumber \\
&&~~~~~~~~~-\Omega\sum_{\nu} g_{\nu\mu}\avg{x(t)f(t)^\nu\left[e^{\frac{-i \widehat x(t)}{\Omega}}-1\right]}_\Upsilon
\EE
This means that the $\widehat R_\mu(t)$ are equal-time response functions (they involve objects such as $\avg{x(t) f(t)\widehat x(t)^\ell}_\Upsilon$), and so by causality we have $\widehat R_\mu(t)=0$.

Finally,
\BE
\frac{\delta}{\delta \widehat R_\mu(t)}\left[\Phi+\Psi+\Upsilon\right]=0 &\Rightarrow&R_\mu(t)=\avg{x(t) f(t)^\mu}_\Upsilon.
\EE
\subsection{Representative species process}
The effective-particle measure that is left therefore reads
\BE
\Upsilon&=&\ln\bigg[ \int Dx D\widehat x Df D\widehat f ~p_0(x)\exp\left(i\sum_{t}\widehat x(t)\left[x(t+\Delta)-x(t)\right]\right)\nonumber \\
&&\times \exp\left(i\Delta\sum_{t} \widehat f(t)f(t)\right) \exp\left(\Omega\Delta\sum_{\mu\nu} g_{\mu\nu} \sum_{t}\left[R_\mu(t) x(t)f(t)^\nu \left[e^{\frac{i\widehat x(t)}{\Omega}}-1\right]+R_\nu(t) x(t) f(t)^\mu \left[e^{\frac{-i\widehat x(t)}{\Omega}}-1\right] \right]\right)\nonumber \\
&& \times\exp\left(-\Delta^2\sum_{t,t} \left[\Gamma K(t',t)x(t)\widehat f(t')+\frac{1}{2}C(t,t')\widehat f(t)\widehat f(t')\right]\right)\bigg].
\EE
With the definition $G(t,t')=-iK(t,t')$ this can be written as
\BE
\Upsilon&=&\ln\bigg[ \int Dx D\widehat x Df D\widehat f~p_0(x) \nonumber \\
&&\exp\left(i\sum_{t}\widehat x(t)\left[x(t+\Delta)-x(t)\right]\right)\nonumber \\
&&\times \exp\left(i\Delta\sum_{t} \widehat f(t)\left[f(t)-\Delta\sum_{t'}\Gamma G(t,t')x(t')\right]-\Delta^2\frac{1}{2}\sum_{i,t,t}   C(t,t')\widehat f(t)\widehat f(t') \right)\bigg]\nonumber \\
&&\times \exp\left(\Omega\Delta\sum_{\mu\nu} g_{\mu\nu} \sum_{t}\left[R_\mu(t) x(t)f(t)^\nu \left[e^{\frac{i\widehat x(t)}{\Omega}}-1\right]+R_\nu(t) x(t) f(t)^\mu \left[e^{\frac{-i\widehat x(t)}{\Omega}}-1\right] \right]\right).
 \EE
 The order parameters are to be determined self-consistently from
 \BE
C(t,t')&=&\avg{x(t)x(t')}_\Upsilon, \nonumber \\
G(t,t')&=&-i\avg{x(t)\widehat f(t')}_\Upsilon,\nonumber \\
R_\mu(t)&=&\avg{x(t) f(t)^\mu}_\Upsilon,
\EE
see above. For any fixed real number $\phi$ we can resum
\BE
\sum_{\mu\nu} g_{\mu\nu} R_\mu(t)\phi^\nu&=&\sum_{\mu\nu} g_{\mu\nu} \phi^\nu \avg{x(t)f^\mu(t)}_\Upsilon \nonumber \\
&=&\avg{x(t) g(f,\phi)}_\Upsilon,
\EE
and similarly
\be
\sum_{\mu\nu} g_{\mu\nu} R_\nu(t)\phi^\mu=\avg{x(t) g(\phi,f)}_\Upsilon.
\ee
We note that the averages on the right are over $x$ and $f$, but not over $\phi$. It is useful to introduce the quantities
\BE
R^+(\cdot,t)&=&\avg{x(t)g[\cdot, f(t)]}_\Upsilon, \nonumber \\
R^-(\cdot,t)&=&\avg{x(t)g[f(t),\cdot]}_\Upsilon,
\EE
where again the average $\avg{\dots}_\Upsilon$ is over $f$ and $x$, but not over the argument of $g$ denoted by $\cdot$. The effective single species measure in final form reads
\BE
\Upsilon&=&\ln\bigg[ \int Dx D\widehat x Df D\widehat f ~p_0(x)\exp\left(i\Delta\sum_{t}\widehat x(t)\frac{x(t+\Delta)-x(t)}{\Delta}\right)\nonumber \\
&&\times \exp\left(\Omega\Delta \sum_{t}\left[x(t)R^-(f,t)\left[e^{\frac{i\widehat x(t)}{\Omega}}-1\right]+x(t)R^+(f,t) \left[e^{\frac{-i\widehat x(t)}{\Omega}}-1\right] \right]\right)\nonumber \\
&&\times \exp\left(i\Delta\sum_{t} \widehat f(t)\left[f(t)-\Delta\sum_{t'}\Gamma G(t,t')x(t')\right]-\Delta^2\frac{1}{2}\sum_{i,t,t}   C(t,t')\widehat f(t)\widehat f(t') \right)\bigg].\label{eq:final}
 \EE
\subsection{Interpretation as birth-death process for a representative species}
In the limit $\Delta\to 0$ the expression in the last line of Eq. (\ref{eq:final}) is recognised as the generating functional of
\be
f(t)=\Gamma\int_{t_0}^t dt' G(t,t') x(t')+\eta(t),\label{eq:fprocess0}
\ee
where $\eta(t)$ is Gaussian noise of zero mean, and with temporal correlations $\avg{\eta(t)\eta(t')}=C(t,t')$. This can be shown by enforcing Eq. (\ref{eq:fprocess0}) via delta-functions in their exponential representation, followed by an average over $\eta$.

The first and second line in Eq. (\ref{eq:final}) on the other hand is the generating functional resulting from a birth-death process, $x\to x\pm\frac{1}{\Omega}$ (equivalent $n\to n\pm 1$) with birth rate $\Omega x(t) R^+[f(t),t]$ and death rate $\Omega x(t) R^-[f(t),t]$. This can be seen by discretising time for such a process, followed by writing down the generating functional for the resulting discrete-time process in which event numbers per time step are Poissonian. One then carries out the average over those Poissonian numbers, in a similar way to the procedure at the beginning of our calculations. Details of the generating functional setup for Markovian and non-Markovian processes can also be found in the Supplement of \cite{brettgalla}.

We conclude the calculation by a brief comment on the quantity $G(t,t')=-i\avg{x(t)\widehat f(t')}_\Upsilon$. Looking at Eqs. (\ref{eq:final}) and (\ref{eq:fprocess0}) (in the limit $\Delta\to 0$) one finds
\be
G(t,t')=\avg{\frac{\delta x(t)}{\delta \eta(t')}}_\star,
\ee
and identifies this quantity as a response function, see also \cite{coolen, sollich, eissopp1}. By causality $G(t,t')=0$ for $t'\geq t$.
\subsection{Gaussian approximation}
The Gaussian approximation is formally obtained by expressing the inner exponentials in Eq. (\ref{eq:final}) as power series in $\Omega^{-1}$, and retaining terms up to and including quadratic order in $\Omega^{-1}$. Within this expansion one has
\BE
\Upsilon&=&\ln\bigg[ \int Dx D\widehat x Df D\widehat f~p_0(x) \exp\left(i\Delta \sum_{t}\widehat x(t)\frac{x(t+\Delta)-x(t)}{\Delta}\right)\nonumber \\
&&\times \exp\left(-i\Delta \sum_{t}\widehat x(t)x(t)[R^+(f,t)-R^-(f,t)] \right) \exp\left(-\frac{1}{2}\frac{\Delta}{\Omega}\sum_{t} x(t)[R^+(f,t)+R^-(f,t)] \widehat x(t)^2 \right)\nonumber \\
&&\times \exp\left(i\Delta\sum_{t} \widehat f(t)\left[f(t)-\Delta\sum_{t'}\Gamma G(t,t')x(t')\right]-\Delta^2\frac{1}{2}\sum_{i,t,t}   C(t,t')\widehat f(t)\widehat f(t') \right)\bigg].\label{eq:final3}
 \EE
 In the limit $\Delta \to 0$ this is seen to describe the process
 \BE
 \dot x &=& x(t)\left[R^+(f,t)-R^-(f,t)\right]+\sqrt{\frac{x(t)\left[R^+(f,t)+R^-(f,t)\right]}{\Omega}}\xi(t), \nonumber\\
 f(t)&=&\Gamma\int^t dt' G(t,t') x(t')+\eta(t),\label{eq:fprocess}
\EE
 with $\xi(t)$ white Gaussian noise of unit variance, and where the above self-consistent relations for the macroscopic order parameters ($R^{\pm}, C, G$) apply. In the main paper we use the notation ${\cal T}^\pm(x)=xR^\pm(f,t)$, and suppress the dependence on $f$ and $t$.
\section{Details of simulation method}
The algorithm generates a set of $M$ sample paths of the effective process \cite{eissopp1,eissopp2}. They are labelled by $\mu=1,\dots,M$. For the purposes of the algorithm we use unit time steps ($t, t+1, \dots$), each such time step corresponds to $\Delta$ units of physical time.
\begin{itemize}
\item[1.] {\em Initialisation.} 

(i) Draw the $x_\mu(0)$ as {\em iid} random variables from an arbitrary initial distribution $p_0(x)$, with non-negative support and with unit mean;

(ii) Compute $C(0,0)=M^{-1}\sum_\mu x_\mu(0)^2$;

(iii) Generate {\em iid} Gaussian random numbers (mean zero, unit variance) $\xi_\mu(0)$, and set $f_\mu(0)=\sqrt{C(0,0)}\xi_\mu(0)$.
\item[2.] {\em Iterate representative process.} 

For all $\mu=1,\dots,M$ carry out the following steps: 

(i) Compute $T_\mu^+=n_\mu\frac{1}{M}\sum_{\nu=1}^M n_\nu g[f_\mu(t),f_\nu(t)]$ and $T_\mu^-=n_\mu\frac{1}{M}\sum_{\nu=1}^M n_\nu g[f_\nu(t),f_\mu(t)]$; 

(ii) Draw Poissonian random numbers $k_\mu^+$ and $k_\mu^-$ with means $T_\mu^+\Delta$ and $T_\mu^-\Delta$; 

(iii) Set $n_\mu(t+1)=n_\mu(t)+k_\mu^+-k_\mu^-$.

\item[3.] {\em Compute order parameters.}

(i) Compute $C(t+1,t')=M^{-1}\sum_\mu x_\mu(t+1)x_\mu(t')$ for all $t'=0,\dots,t+1$; 

(ii) Compute the vector $\mathbf{b}=(b_0,\dots,b_{t+1})$, where $b_{t'}=M^{-1}\sum_\mu \eta_\mu(t')x_\mu(t+1)$;

(iii) Solve the system $\mathbf{C}\mathbf{g}=\mathbf{b}$ to obtain $\mathbf{g}=(G_{t+1,0},\dots,G_{t+1,t+1})$. The upper left $(t+1)\times (t+1)$ block of the response matrix is now available.

\item[4.] {\em Generate Gaussian noise and compute fitness.}

(i) Carry out a Cholesky decomposition of $\mathbf{C}$, i.e. find a lower triangular matrix $\mathbf{B}$ such that $\mathbf{C}=\mathbf{B}^T\mathbf{B}$; 

(ii) Draw $\xi_\mu(t+1)$ as {\em iid} standard Gaussians ($\mu=1,\dots, M$), and for each $\mu$ set $\eta_\mu(t+1)=\sum_{t'} B_{t+1,t'}\xi_\mu(t')$; 

(iii) For each $\mu$ calculate $f_\mu(t+1)=\Gamma\sum_{t'=0}^{t+1}G(t+1,t') x_\mu(t')+\eta_\mu(t+1)$.
\item[5.] Goto 2. and iterate.
\end{itemize}

\end{appendix}


\begin{thebibliography}{10}
\bibitem{remark} We plot $n_i=x_i\Omega$, where $\Omega=50$ and with $x_i$ obtained from Eq. (\ref{eq:det}).
\bibitem{note} The distribution of the variables $k_{ij}(t)$ will depend on the $\{n_i\}$ up to time $t$. The sum over the $\bn$ and $\bk$ in the generating functional (\ref{eq:genf}) has to be carried out iteratively; see also \cite{brettgalla}.
\bibitem{remark1} The first bin in the histogram of the inset is necessarily populated, even if none of $x_i$ has strictly reached zero at finite times under the deterministic dynamics.
\end{thebibliography}
\end{document}